\pdfoutput=1
\documentclass[aps,prl,twocolumn,showpacs,superscriptaddress]{revtex4}
\usepackage{graphicx}
\usepackage{epstopdf}
\usepackage{latexsym}
\usepackage{amssymb}
\usepackage{amsmath}
\usepackage{amsfonts}
\usepackage{subfigure}
\usepackage{bm}
\usepackage{multirow}

\newcommand{\ket}[1]{|#1 \rangle}

\newcommand{\ZZ}{\mathbb{Z}}

\newcommand{\eqnref}[1]{Eq.\,\eqref{#1}}
\newcommand{\figref}[1]{Fig.\,\ref{#1}}

\begin{document}

\title{Projective non-Abelian Statistics of Dislocation Defects in a $\ZZ_N$ Rotor Model}
\author{Yi-Zhuang You}
\affiliation{Institute for Advanced Study, Tsinghua University, Beijing, 100084, China}
\author{Xiao-Gang Wen}
\affiliation{Department of Physics, Massachusetts Institute of Technology, Cambridge, Massachusetts 02139, USA}
\affiliation{Perimeter Institute for Theoretical Physics, Waterloo, Ontario, N2L 2Y5 Canada}

\date{\today }

\begin{abstract}
Non-Abelian statistics is a phenomenon of topologically protected non-Abelian Berry phases as we exchange \emph{quasiparticle excitations}.  In this paper, we construct a $\ZZ_N$ rotor model that realizes a self-dual $\ZZ_N$ Abelian gauge theory.  We find that lattice dislocation defects in the model produce topologically protected degeneracy. Even though dislocations are not quasiparticle excitations, they resemble non-Abelian anyons with quantum dimension $\sqrt{N}$.  Exchanging dislocations can produces topologically protected projective non-Abelian Berry phases.  The dislocations, as projective non-Abelian anyons can be viewed as a generalization of the Majorana zero modes.
\end{abstract}

\pacs{05.30.Pr, 05.50.+q, 61.72.Lk, 03.67.-a}
\maketitle

\emph{Introduction}--- Searching for Majorana fermions (or more precisely, Majorana zero modes) in condensed matter systems have attracted increasing research interests recently.\cite{Wilczek, ReadGreen, Ivanov, DasSarma1, He3, FuKane, DasSarma2, DasSarma3, Alicea} But what really is the Majorana zero mode?  In fact, the so called  ``Majorana zero mode'' is actually a phenomenon of topologically protected degeneracy in the presence of certain topological defects (such as vortices in 2D $p_x+ip_y$ superconductors\cite{ReadGreen, Ivanov}). In the race for finding Majorana zero modes, much attention has been paid to the fermion systems.\cite{DasSarma1, He3, FuKane, DasSarma2, DasSarma3, Alicea} However the boson/spin systems also have topologically protected degeneracies,\cite{Wtop,Wrig,ReadSachdev, WenZ2, Moessner, Kitaev, KitaevHC, WenPlaquette, LevinWen1, LevinWen2} which may also be ascribed to Majorana zero modes or their generalizations.

An 1D example of emergent Majorana zero modes in the spin system arises from the transverse field Ising chain,\cite{K0131,Sachdev} whose ground state degeneracy in the ferromagnetic phase can be viewed as the Majorana zero modes at both ends of the chain. A 2D example is found in the toric code model\cite{Kitaev, WenPlaquette}, where lattice dislocations are braided and fused as if they were Majorana zero modes\cite{Bombin, KitaevKong} which resemble non-Abelian anyons\cite{MooreRead,Wnab,RMP} of quantum dimension $\sqrt 2$.  The toric code model can be generalized to a $\ZZ_N$ rotor model, whose low energy effective theory is a self-dual $\ZZ_N$ gauge theory.\cite{ZN1,ZN2} In this paper, we study the topologically protected degeneracy associated with the extrinsic topological defects, namely lattice dislocations in the $\ZZ_N$ rotor model, and found that these defects are of quantum dimension $\sqrt{N}$, which can be viewed as a generalization of the ``Majorana zero mode''. Braiding topological defects with protected degeneracy will lead to topologically protected projective non-Abelian Berry phase, which may allow us to perform decoherence free quantum computations.\cite{RMP} 

We like to remark that the dislocations in our $\ZZ_N$ rotor model are \emph{not} non-Abelian anyons, since the non-Abelian anyons must be excitations of the Hamiltonian, while the dislocations are not the excitations in this sense. The dislocations do not really carry non-Abelian statistics since the non-Abelian Berry phase from exchanging dislocations is topologically protected only up to a total phase.  We say the dislocations carry a projective non-Abelian statistics.\cite{pST1,pST2} An other example of projective non-Abelian statistics for dislocations in fractional quantum Hall states on lattice can be found in Ref. \onlinecite{BQ}.

\begin{figure}[b]
\begin{center}
\includegraphics[height=0.084\textheight]{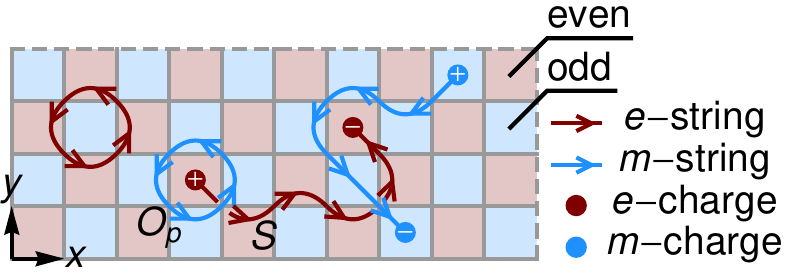}
\end{center}
\caption{(Color on line.) Even$\times$even periodic lattice with plaquettes colored in a check board pattern: red and darker (blue and lighter) plaquette will be called even (odd). Each directed string represent a product of $U_i$ and/or $V_i$ operators on the sites along the string. The operator on each site is specified by the string direction (see text).}
\label{fig:lattice}
\end{figure}

\emph{$\ZZ_N$ plaquette model}--- The $\ZZ_N$ plaquette model is a rotor model on a two-dimensional square lattice (see \figref{fig:lattice}). On each site $i$, define a $\ZZ_N$ rotor with $N$ basis states $\ket{m_i}$, labeled by the angular momentum $m_i=0,1,\cdots,(N-1)$. For each rotor, introduce $U_i$ to measure the angular momentum $U_i\ket{m_i}=e^{i \theta_N m_i}\ket{m_i}$ with $\theta_N\equiv2\pi/N$, and $V_i$ to lower the angular momentum by one $V_i\ket{m_i}=\ket{(m_i-1)_\text{mod $N$}}$. Both $U_i$ and $V_i$ are unitary operators $U_i^\dagger U_i=V_i^\dagger V_i=1$,
satisfying $V_i U_{i'}= e^{i\theta_N\delta_{ii'}}U_{i'} V_i$.

The $\ZZ_N$ plaquette model is given by the Hamiltonian
\begin{equation}\label{eq:H}
H= - \sum_p O_p +h.c.,
\end{equation}
where the operator $O_p$ describes a kind of ring coupling among the rotors on the corner sites of each plaquette $p$,
\begin{equation}\label{eq:O4}
O_p = \vcenter{\hbox{\includegraphics[height=24pt]{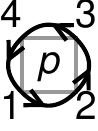}}} = U_1 V_2 U_3^\dagger V_4^\dagger.
\end{equation}
Here we adopt the graphical representation for the operators: $U_i=\vcenter{\hbox{\includegraphics[height=8pt]{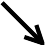}}}$, $V_i=\vcenter{\hbox{\includegraphics[height=8pt]{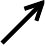}}}$, $U_i^\dagger=\vcenter{\hbox{\includegraphics[height=8pt]{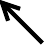}}}$, $V_i^\dagger=\vcenter{\hbox{\includegraphics[height=8pt]{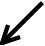}}}$, by drawing directed strings going through the site. Because these operators only connect diagonal plaquettes, a string starting from the even plaquette will never enter the odd plaquette (and vice versa). So we can locally distinguish two different types of strings: $e$-string ($m$-string) if it lives in the even (odd) plaquettes (see \figref{fig:lattice}). The assignment of even/odd to the plaquettes can be reversed under the translation of one lattice spacing, so the interchange of $e$- and $m$-strings could be realized by the curvature of the lattice as will be seen later.

The $\ZZ_N$ plaquette model \eqnref{eq:H} is exact solvable, as evidenced from the commutation relation $[O_p,O_{p'}]=0$, as $O_p O_{p'}=\vcenter{\hbox{\includegraphics[height=16pt]{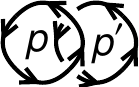}}}=e^{i\theta_N}e^{-i\theta_N}\vcenter{\hbox{\includegraphics[height=16pt]{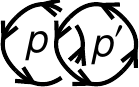}}}=O_{p'}O_p$ for adjacent $p$ and $p'$, where the overlay of strings indicates the ordering of the operators, such as $V_iU_i = \vcenter{\hbox{\includegraphics[height=9pt]{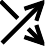}}}$ and $U_iV_i=\vcenter{\hbox{\includegraphics[height=9pt]{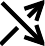}}}$, with the algebra $\vcenter{\hbox{\includegraphics[height=9pt]{fig_VU.pdf}}}=e^{i\theta_N}\vcenter{\hbox{\includegraphics[height=9pt]{fig_UV.pdf}}}$.

Every $O_p$ operator has $N$ distinct eigenvalues $e^{i\theta_N q_p}$ labeled by $q_p=0,1,\cdots,(N-1)$, as inferred from the fact that $O_p^N\equiv 1$. $q_p$ denotes the (generalized) $\ZZ_N$ charge hosted by the plaquette $p$. If the plaquette is even (odd), we may call it $e$-charge ($m$-charge). The energy  will be minimized if all $O_p$'s take the eigenvalue 1 ($q_p=0$). Therefore the ground states are the common eigenstates that satisfy $O_p\ket{\text{grnd}}=\ket{\text{grnd}}$ for all $p$'s, and is free of any $\ZZ_N$ charges.

\emph{Intrinsic anyon excitations}--- The excited states can be obtained by applying open string operators to the ground state, which create opposite $\ZZ_N$ charge excitations in pairs at both ends of the string. These excitations and can be detected by the close string operator (like $O_p$) surrounding them in the counterclockwise direction. Take $S$ in \figref{fig:lattice} for example, $ O_p S\ket{\text{grnd}}=\vcenter{\hbox{\includegraphics[height=16pt]{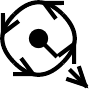}}}\cdots\ket{\text{grnd}}= e^{i\theta_N}\cdots \vcenter{\hbox{\includegraphics[height=16pt]{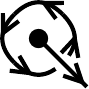}}}\ket{\text{grnd}}= e^{i\theta_N}S\ket{\text{grnd}}$, showing that a charge $q_p=+1$ is created at the end of the open string by the action of $S$. One can show that the
opposite charge $q_p=-1$ is created at the other end.

Because $\ZZ_N$ charge excitations are the ends of open strings, their statistics are inherited from the algebra of the string operators. According to $\vcenter{\hbox{\includegraphics[height=10pt]{fig_VU.pdf}}}=e^{i\theta_N}\vcenter{\hbox{\includegraphics[height=10pt]{fig_UV.pdf}}}$, braiding a $q_{e}$ $e$-charge with a $q_{m}$ $m$-charge would acquire a phase $e^{i(\theta_N/2)q_{e}q_{m}}$. In this sense, these excitations are Abelian anyons. However we must stress that these anyons are \emph{intrinsic}, as they are collective motions of rotors, described by the excited state within the rotor Hilbert space. This is to be distinguished from the \emph{extrinsic} anyons introduced later as lattice dislocations, which does not belongs to the rotor Hilbert space. Note that both the phase $e^{i(\theta_N/2)q_{e}q_{m}}$ and the excitation energy are invariant under the exchange of $e$ and $m$. This manifests the self-duality of the $\ZZ_N$ plaquette model, and can be realized by lattice translation.

\emph{Ground state degeneracy}--- The degeneracy of the ground states of the $\ZZ_N$ plaquette model depends on the topology of the lattice. Let us consider the torus topology by setting the model on a $L_x\times L_y$ sized lattice with periodic boundary condition in both directions. The total number of states is $N^{N_\text{site}}$, with $N_\text{site}=L_xL_y$ being the number of sites. To count the ground states, we note that they are constrained by $\forall p: O_p=1$.  Consider a particular $O_p$ operator and the subspaces labeled by its different eigenvalues. Those subspaces all have the same dimension, because any open string operator that ends in the plaquette $p$ can be used to perform a unitary transform that rotates these subspaces into each other. So each time imposing $O_p=1$ on a particular plaquette will reduce the available Hilbert space dimension by a factor of $N$.  However the $O_p$ operators are not independent. Because $e$-charges ($m$-charges) are created in opposite pairs, summing over the lattice, $e$-charges and $m$-charges must be neutralized respectively, i.e. $\prod_{p\in\text{even}}O_p=\prod_{p\in\text{odd}}O_p=1$. This is true on an even$\times$even lattice (i.e. both $L_x$, $L_y$ are even), which reduces the number of independent $O_p$ constrains to $(N_\text{plaq}-2)$, with $N_\text{plaq}=L_xL_y$ being the number of plaquettes. So after restricting the full Hilbert space to the ground state subspace, the remaining dimension is $N^{N_\text{site}-N_\text{plaq}+2}=N^2$, meaning the ground state degeneracy of the $\ZZ_N$ plaquette model is $N^2$ on the even$\times$even lattice. However for the even$\times$odd or odd$\times$odd lattices (i.e. $L_x$ or $L_y$ is odd), $e$-string and $m$-string can be continued into each other by going along the odd direction, thus $e$-charge and $m$-charge are made identical. So they are no longer required to be neutralized respectively, but only neutralized as a whole. Therefore we only have one relation $\prod_p O_p=1$, which reduces the number of independent $O_p$ constraints to $(N_\text{plaq}-1)$, and the resulting ground state degeneracy will be $N^{N_\text{site}-N_\text{plaq}+1}=N$.

To summarize, the ground state degeneracy of the $\ZZ_N$ plaquette model on a torus follows from the general formula
\begin{equation}\label{eq:GSD}
\text{GSD} = \mathcal{N} N^{N_\text{site}-N_\text{plaq}},
\end{equation}
where $\mathcal{N}$ denotes the number of species of the intrinsic excitations that are supported by the lattice topology. On the even$\times$even lattice, we have totally $\mathcal{N}=N^2$ distinct excitations by combination of $e$- and $m$-charges. When it comes to the even$\times$odd or odd$\times$odd lattice, $e$- and $m$-charges are no longer distinct, and the number of excitation species is reduced to $\mathcal{N}=N$. The topological order in the ground state is now evidenced from the protected ground state degeneracy on torus,\cite{Wtop,Wrig} and from the dependence of the  ground state degeneracy on the parity of the lattice periodicity.

\begin{figure}[tb]
\begin{center}
\subfigure[]{\includegraphics[height=0.125\textheight]{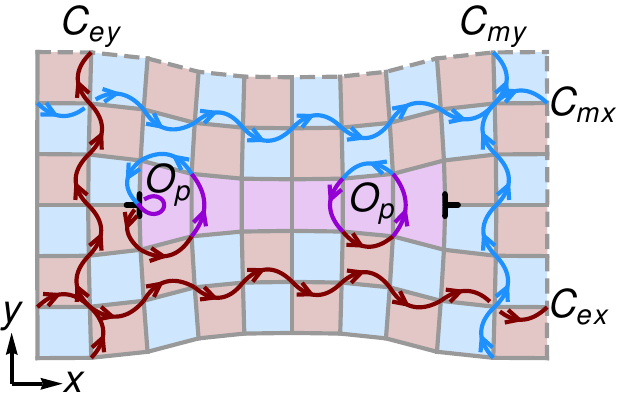}\label{fig:dislocation}}
\subfigure[]{\includegraphics[height=0.066\textheight]{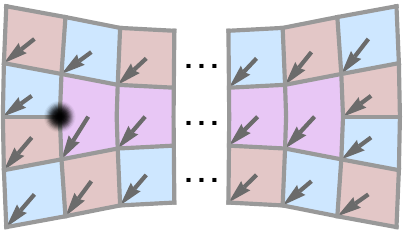}\label{fig:p2s}}
\end{center}
\caption{(Color on line.) \subref{fig:dislocation} Lattice with a pair of dislocations, marked out by ${\pmb{\dashv}}$ and $\pmb{\vdash}$. Plaquettes on the branch cut are colored by violet. Periodic boundary conditions are assumed in both direction by sticking the dashed edges with the solid edges on the opposite side. The ring operators $O_p$ are redefined around the pentagonal plaquette. $C$ operators denote large close strings looping around the lattice. \subref{fig:p2s} Plaquette to site mapping. The site that is not mapped to is marked by a black dot.}
\end{figure}

\emph{Dislocations}--- One can change the lattice periodicity by first generating a pair of edge dislocations with opposite unit length Burger's vectors, and moving them in the direction perpendicular to their Burger's vectors all the way around the lattice, then annihilating them as they meet again at the periodic boundary. During this process, the ground state degeneracy must have changed. This motivates us to introduce \emph{dislocations} as shown in \figref{fig:dislocation} to probe the topological order by looking at the degeneracy associated to them. With dislocations, one can no longer globally color the plaquettes consistently. \emph{Branch cuts} must be left behind between each pairs of dislocations. Going around a dislocation exchange the $e$- and $m$-charges, as $e$- and $m$-strings are transmuted into each other across the branch cut. The self-duality is made explicit by dislocations. 

In the presence of dislocations, the $\ZZ_N$ plaquette model is still defined by the Hamiltonian in \eqnref{eq:H}, with the same ring operator $O_p$ in \eqnref{eq:O4} for quadrangular plaquettes (including those on the branch cuts). Only around the pentagonal plaquettes (at the dislocations), the ring operator $O_p$ should be redefined as
\begin{equation}
O_p =  -e^{i\frac{\theta_N}{2}}\;\vcenter{\hbox{\includegraphics[height=32pt]{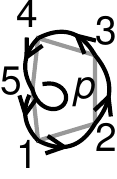}}} = -e^{i\frac{\theta_N}{2}}  U_1 V_2 U_3^\dagger V_4^\dagger U_5 V_5^\dagger.
\end{equation}
The phase factor $-e^{i\theta_N/2}$ is to guarantee that $O_p^N\equiv 1$ holds for the pentagonal plaquette as well. The pentagonal ring operator $O_p$ commutes with all the other ring operators, so the exact solvability of the model is preserved. The ground states are again common eigenstates of $\forall p:O_p\ket{\text{grnd}}=\ket{\text{grnd}}$. The dislocations are topological defects that do not belong to the model Hilbert space. To distinguish from those \emph{intrinsic} $\ZZ_N$ charges, we will call the dislocations as the \emph{extrinsic} defects.

With the branch cuts, $e$-charge and $m$-charge are indistinguishable, so the species of intrinsic excitations count to $\mathcal{N}=N$. According to \eqnref{eq:GSD}, the ground state degeneracy will be given by $N^{N_\text{site}-N_\text{plaq}+1}$ in general. To count the number of sites and plaquettes, we first establish a correspondence between them by mapping each plaquette to its bottom-left corner site, as indicated by the arrows in \figref{fig:p2s}. Between a pair of dislocations, only one of them will hold a site that has no plaquette correspondence (see \figref{fig:p2s}), so the introduction of every pair of dislocations will give rise to one extra site (with respect to the number of plaquettes). Therefore if there are $n$ dislocations on the lattice, there will be $N_\text{site}-N_\text{plaq}=n/2$ more sites than plaquettes, and the ground state degeneracy of the $\ZZ_N$ plaquette model will be $\text{GSD}=N^{n/2+1}$.

\begin{figure}[tb]
\begin{center}
\includegraphics[height=0.15\textheight]{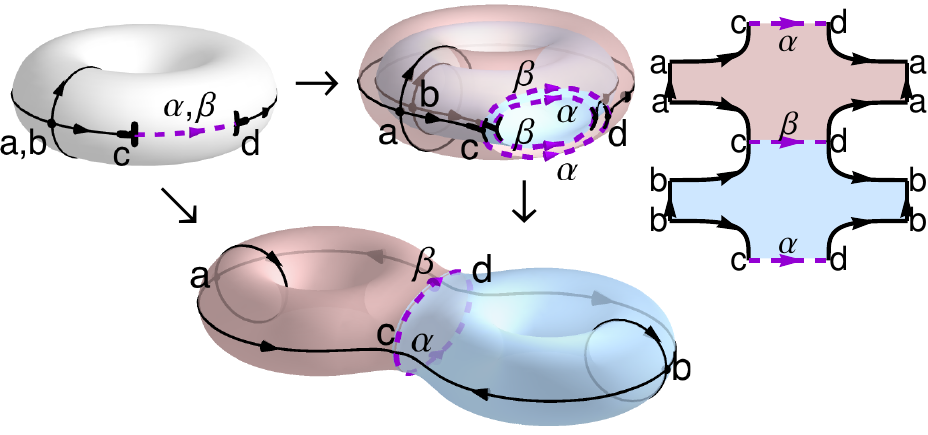}
\end{center}
\caption{(Color on line.) Diffeomorphism of the torus with a pair of dislocations at $c$ and $d$. Expand the branch cut (violet dashed line) between $c$ and $d$ into a hole, with two edges marked by $\alpha$ and $\beta$. Separate the $e$- and $m$-layers. Unwrap both layers by cutting along the large loops around the torus. Rotate one layer to glue the $\beta$ edges together along the marked direction. Glue the other edges and rewrap into a double torus.}
\label{fig:Tcut}
\end{figure}

This ground state degeneracy is topologically protected indeed. To better understand the topology, we start from the even$\times$even periodic lattice without dislocations, i.e. a torus with no branch cut. In this case, the $e$-strings and $m$-strings are distinct, and can never be deformed into each other, as if they were living on two different layers of the torus. So the topological space is the disjoint union of two separate torus. Introducing a pair of dislocations, the two layers will be connected: strings on one layer can be carried on into the other layer through the branch cut. So the topological space becomes a doubled torus under the diffeomorphism\cite{BarkeshliWen} as shown in \figref{fig:Tcut}.

\begin{figure}[tb]
\begin{center}
\subfigure[]{\includegraphics[width=0.4\textwidth]{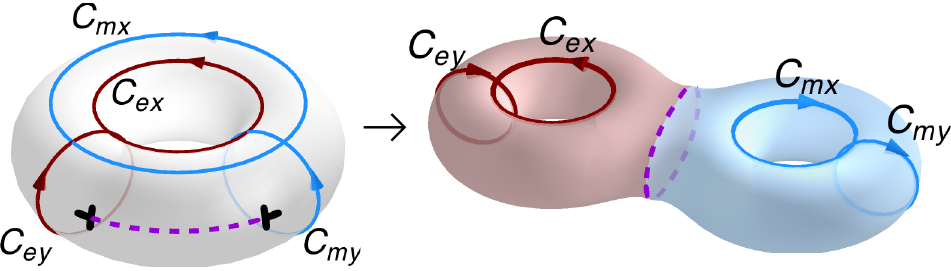}\label{fig:T2}}\quad
\subfigure[]{\includegraphics[width=0.4\textwidth]{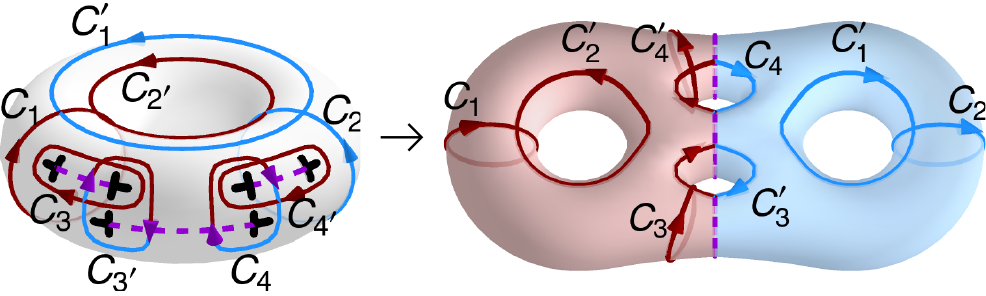}\label{fig:T4}}
\end{center}
\caption{(Color on line.) Diffeomorphism of string operators on the torus \subref{fig:T2} with a pair of dislocations, or \subref{fig:T4} with 3 pairs of dislocations.}
\end{figure}

All the operators that act within the ground state subspace are closed-string (cycle) operators, as they commute with the Hamiltonian.  Note that the contractable cycles act trivially (as $O_p=1$). Only non-contractable cycles can be used to label the different ground states and to perform unitary transforms among them. On the double torus topology as in \figref{fig:T2}, one can specify 4 non-contractable cycles: $C_{ex}$, $C_{ey}$, $C_{mx}$, $C_{my}$, as the canonical homology basis. Their operator forms are given explicitly according to their graphical representations depicted in \figref{fig:dislocation}. We now study the representation of these cycle operators in the ground state subspace. First we find the following commutation relations  $[C_{ex},C_{ey}] = [C_{mx},C_{my}] = [C_{ex},C_{mx}] = [C_{ey},C_{my}] = 0$, and two independent algebras $C_{ey}C_{mx}=e^{i\theta_N}C_{mx}C_{ey}$, $C_{my}C_{ex}=e^{i\theta_N} C_{ex}C_{my}$. Each algebra requires an $N$-dimensional representation space, so the 4 cycle operators together requires $N^2$-dimensional representation space, which must have completed the ground state subspace, since all the non-contractable cycles can be generated by these 4 basis cycles. Therefore the ground states are $N^2$-fold degenerated, and each of them corresponds to a basis in the representation space. Any perturbation of the Hamiltonian that is non-zero only in a compact region will not change the ground state degeneracy, since the non-contractable cycle operators that avoid the compact region still commute with the Hamiltonian.

The above can be generalized to the case with any number of dislocations. Consider $n$ dislocations with $n/2$ branch cuts. Following the similar cut-and-glue procedures in \figref{fig:Tcut}, the topological space will be a genus $g=n/2+1$ surface as in \figref{fig:T4}, on which one can choose $g$ pairs of non-contractable cycle operators $C_a$ and $C'_a$ ($a=1,\cdots,g$), such that $[C_a,C_b]=[C'_a,C'_b]=0$ and $C_aC'_b=e^{i\theta_N\delta_{ab}}C'_bC_a$. These operators spans a $N^g$-dimensional representation space isomorphic to the ground state subspace. Therefore the ground state degeneracy of the $\ZZ_N$ plaquette model with $n$ dislocations is $\text{GSD}=N^{n/2+1}$, which is consistent with our previous result. Each dislocation contributes to the ground state degeneracy by a factor of $\sqrt{N}$. Thus the dislocations resemble non-Abelian anyons of quantum dimension $\sqrt{N}$, as described in Ref.\,\onlinecite{BarkeshliWen}. Braiding the dislocations leads to topologically protected projective non-Abelian Berry phases.  We see that projective non-Abelian anyon can emerge from an Abelian model as the extrinsic topological defects, such as lattice dislocations. Those projective non-Abelian anyon can be used to perform topological quantum computations,\cite{RMP} but not universally since the square of the quantum dimension is an integer.\cite{RSW}

\emph{Parton approach}--- For the $N=2$ case, the quantum dimension $\sqrt{2}$ implies that the extrinsic anyons are Majorana fermions. To expose the Majorana fermion explicitly, we evoke the parton projective construction, in which 4 Majorana fermions $\eta_i^\alpha$ ($\alpha=1,2,3,4$) are introduced on each site $i$, obeying the anti-commutation relation $\{\eta_i^\alpha,\eta_j^\beta\}=\delta_{ij}\delta_{\alpha\beta}$.\cite{WenPlaquette} Under the constraint $\eta_i^1\eta_i^2\eta_i^3\eta_i^4=1/4$, the rotor operators can be expressed as $U_i=i\eta_i^1\eta_i^2$, $V_i=i\eta_i^2\eta_i^3$. Then the $\ZZ_2$ plaquette model can be mapped to an interacting fermion model, which has a ``mean-field'' description given by $H_\text{mean}=-\sum_{\langle ij\rangle}(s_{ij}\Delta_{ij}+h.c.)$ with the ansatz $s_{ij}=\pm1$ on each bound, where $\Delta_{i,i+\hat{x}}=i\eta_i^1\eta_{i+\hat{x}}^3$ and $\Delta_{i,i+\hat{y}}=i\eta_i^2\eta_{i+\hat{y}}^4$. Let $\ket{\{s_{ij}\}}$ be a free fermion ground state of $H_\text{mean}$, and $\mathcal{P}=\prod_i\frac{1}{2}(1+4\eta_i^1\eta_i^2\eta_i^3\eta_i^4)$ be the projection operator to the physical Hilbert space of rotors. All the eigenstates of the $\ZZ_N$ plaquette model can be obtained by the projective construction as $\mathcal{P}\ket{\{s_{ij}\}}$. To obtain the ground states, $\{s_{ij}\}$ must satisfy the flux configuration given by $O_p=1$, which has totally 4 gauge inequivalent solutions on a torus. Given a particular $\{s_{ij}\}$, all the Majorana fermions will be paired up across the bound, except for the dangling Majorana fermion at the dislocation site. If there are $n$ dislocations in the system, there will be $n$ dangling Majorana zero modes, which leads to a $2^{n/2}$ fold degeneracy in the free fermion ground states. So altogether we have $4\times2^{n/2}$ fermion states to be projected from, half of which will be projected to nothing due to their odd fermion parity. Therefore the resulting physical ground states add up to $4\times2^{n/2}/2=2^{n/2+1}$, consistent with our previous formula. The above discussion has shown that the $\sqrt{2}$ quantum dimension of the extrinsic anyon actually originated from the dangling Majorana fermion, or the Majorana zero mode, at the dislocation site. It has been shown that exchanging  Majorana zero modes will lead to non-Abelian Berry phase, which supports our conjecture that exchanging dislocations in our $\ZZ_N$ plaquette model leads to protected (projective) non-Abelian Berry phase.

In conclusion, we studied the phenomenon of topologically protected degeneracy and  topologically protected projective non-Abelian Berry phases produced by extrinsic topological defects (such as dislocations) in a $\ZZ_N$ rotor model. We find that these dislocations are projective non-Abelian anyons with quantum dimension $\sqrt{N}$. For $N=2$, such a result can be re-derived from a parton construction where the dislocations can be identified as Majorana zero modes. For higher $N$ ($N>2$), the projective non-Abelian anyons ({\it i.e.} the dislocations) can be viewed as a generalization of the Majorana zero modes.

\begin{acknowledgments}
We would like to thank Zhenghan Wang, Zheng-Cheng Gu, and Liang Kong for helpful discussions.  This work is supported by NSF Grant No. DMR-1005541 and NSFC 11074140.
\end{acknowledgments}

\end{document}